\documentclass[aps,twocolumn,prl]{revtex4}
\usepackage{color}

\usepackage{epsfig}
\usepackage{epstopdf}
\usepackage{amsmath}
\usepackage[latin1]{inputenc}
\usepackage{graphicx}

\newcommand{\nc}{\newcommand}
 \nc{\tcb}{\textcolor{blue}}  
 \nc{\tcr}{\textcolor{red}}
 \nc{\be}{\begin{equation}} 
 \nc{\ee}{\end{equation}}
 \nc{\bt}{\begin{tabular}} 
 \nc{\et}{\end{tabular}}
 \nc{\bea}{\begin{eqnarray}}  
 \nc{\eea}{\end{eqnarray}}
 \nc{\ba}{\begin{array}}  
 \nc{\ea}{\end{array}}
 \nc{\rds}{{\rm d}s} 
 \nc{\rdt}{{\rm d}t} 
 \nc{\rdr}{{\rm d}r}
 \nc{\rdO}{{\rm d}\Omega} 
 \nc{\s}{{\rm S}} 
 \nc{\Pl}{{\rm Planck}}
 \nc{\dis}{\displaystyle} 
 \nc{\crit}{_{\rm cr}} 
 \nc{\rd}{{\rm d}}
 \nc{\munu}{{\mu\nu}} 
 \nc{\erm}{{\rm e}}
 \nc{\drm}{{\rm d}}
 \nc{\ov}{\overline}

\begin{document}

\title{Minimum light transmission in graphene in the presence of a magnetic field}

\author{S. Esposito}
\email{Salvatore.Esposito@na.infn.it}%
\affiliation{
Istituto Nazionale di Fisica Nucleare, Sezione di Napoli, Complesso Universitario di Monte
S.\,Angelo, via Cinthia, I-80126 Naples, Italy}

\

\

\begin{abstract}
\noindent We show that, on general theoretical grounds, transmission of light in graphene always presents a non-vanishing minimum value independently of any material and physical condition, the transmission coefficient being higher in the presence of a substrate, and getting increasing when QED corrections higher than $\alpha$ come into play. Explicit numerical calculations for typical cases are carried out when an external magnetic field is applied to the sample, showing that, in epitaxial graphene, a threshold effect exists leading to a non trivial minimum transmission, for a non vanishing light frequency, only for field values larger than a critical one, both in the large and in the intermediate chemical potential regime. Such a threshold effect manifests even in the maximum Faraday rotation polarization of light, which is substantially controlled by the applied magnetic field. Instead, more transmission minima in suspended graphene enters in the considered light frequency region for increasing magnetic field, displaying an effective shift of frequency bands where the sample gets more or less absorptive with a suitable tuning of the external field. Two transition regions in different magnetic field ranges are found, where the shift effect towards higher frequency values occurs both in the transmission coefficient and in the Faraday rotation angle. Potential technological application of the results presented are envisaged.


%Keywords: light transmission in graphene; magnetized graphene; threshold effects in graphene

\end{abstract}

\maketitle

\noindent The naturally-occurring single sheet of carbon atoms in graphene \cite{Novo2005} has attracted considerable interest in the last decade in different areas of research and technology \cite{Geim2007}-\cite{Sarma2011}, due to the peculiar and particularly intriguing properties of such 2D material. Its hardness, yet flexibility, indeed, as well as high electron mobility and thermal conductivity, has put graphene in the spotlight of applied research in condensed matter physics. Also, the simple theoretical description \cite{Gusynin2005} -- confirmed by experimental evidences \cite{Zhang2005} -- has showed that the properties of charge carriers in graphene are completely similar to those of ultrarelativistic electrons \cite{Castro2009}, the corresponding quasiparticles obeying a linear dispersion relation, so that a new era of Dirac materials has opened with potential applications in nanotechnology. 

Particularly extraordinary are the optical properties of monolayer graphene which, despite being only one atom thick, presents a surprisingly huge effect of absorption of a significant 2.3\% fraction of the incident light \cite{Nair2008}, as a consequence of its unique conical electronic band structure \cite{Castro2009}. This is just the behavior expected for ideal Dirac fermions \cite{Kuz2008}-\cite{Stauber2008}, and the result proved to be valid for a wide range of frequencies. Graphene's opacity $1-T \simeq \pi \alpha \simeq 2.3 \%$ can be indeed obtained by calculating the absorption of light by two-dimensional Dirac particles with Fermi's golden rule, and its only dependence on the fine structure function $\alpha$ is a consequence of the fact that the optical conductivity $\sigma = \pi e^2/ 2 h$ ($e$ and $h$ being the electron charge and the Planck constant, respectively) is independent of any material parameter.

Surprising results are coming also from magneto-optical experiments in mono and multilayer graphene \cite{Sado2006}-\cite{Grassee2011}, including the unexpectedly large Faraday rotation effect, which motivated both further experimental searches and theoretical studies \cite{Gusynin2007}-\cite{Vale2015}.

In general, the possibility to tune the optical properties of graphene by means of material or external parameters would speed up further its applications in different areas of technology. For example, since single-layer graphene is transparent to light to a high degree,  it proves to be very promising as a protection layer for optical devices as mirrors, lenses and screens. On the other hand, however, full or near full light absorption \cite{Apell2012}-\cite{Hashe2013} achieved, for instance, by controlling the chemical potential with a volta\-ge bias and/or doping, or by using patterned metallic nanostructures, opens interesting possibilities in ultra-fast optoelectronic applications, graphene-based photovoltaics and, more in general, in boosting the efficiency of THz and infrared detection. 

Irrespective of the given final application, here we focus on the problem of maximum absorption controlled simply -- and mainly -- by an external magnetic field, where light transmission in suspended and epitaxial graphene is calculated, as well as the Faraday rotation effect.

Tight-binding approach to the description of monolayer graphene has proved quite successful, and since it is equivalent, for small momenta, to the relativistic Dirac model of quasiparticles (with the speed of light replaced by $v_F \simeq c/300$), we definitely adopt such a model in our calculations (with the  caveat that parameters of the Dirac model may differ from sample to sample). The Dirac Hamiltonian describing the system is then $(a=1,2)$:
\be \label{dh}
H = - i v_F \gamma^0 \gamma^a \partial_a \, , 
\ee
\[ \footnotesize
\gamma^0 = \left( \ba{cc} \sigma_3 & 0 \\ 0 & \sigma_3 \ea \right) \! , \quad
\gamma^1 = i \left( \ba{cc} \sigma_1 & 0 \\ 0 & \sigma_1 \ea \right) \! , \quad
\gamma^2 = i \left( \ba{cc}  \sigma_2 & 0 \\ 0 & - \sigma_2 \ea \right) \! , 
\]
${}$ \\
where the pseudo-spin index refers to the sublattice degree of freedom. As in some recent literature \cite{Gusynin2007, Fialko2012, Vale2015}, our magneto-optical calculations will adopt the language of quantum field theory, which is certainly more adequate to describe graphene properties than non-relativistic quantum mechanics, since the tight-binding model corresponds, in the continuum limit, to massless quantum electrodynamics in (2+1) dimensions, with a static Coulomb interaction varying as the inverse of the distance, as in ordinary space \cite{Gusy2007}. 

Let us consider the scattering of light by an infinite graphene membrane immersed in a (3+1)-dimensional ambient space oriented in the $xy$ plane and placed in an external electromagnetic field of potential $A$. The interaction with quasi-particles propagating in the graphene surface is described by the Dirac action ($j=0,1,2$)
\be \label{da}
S = \int \drm^3 x \, \ov{\psi} {\not D} \psi \, , \qquad {\not D} = i \tilde{\gamma}^j \left( \partial_j + i e A_j \right) 
\ee
($\tilde{\gamma}^0 = \gamma^0, \tilde{\gamma}^{1,2} = v_F \gamma^{1,2}$). Note that, while the spinors are confined to the graphene surface, the electromagnetic field lives in the ambient (3+1)-dimensional space. From this action it is then straightforward, according to the standard quantum field theory formalism, to evaluate the polarization operator $\Pi_{ab}$ entering in the expression for the ac conductivity $\sigma_{ab} = \Pi_{ab}/i \omega$, from which the optical properties of graphene directly follows. With the same notation of Ref. \cite{Fialko2012}, the total transmission $T$ of linearly polarized (along the $x$-axis) light of frequency $\omega$, passing normally through the graphene layer on a substrate of a finite thickness $d$ and refractive index $n_s$, is written as:
\be \label{tt}
T = 16 n_s^2 \, \frac{|A_{x}|^2 + |A_{y}|^2}{|A_{x}^2 + A_{y}^2|^2} \, ,
\ee
\bea
A_x &=& (n_s -1)(n_s-1-\sigma_{xx}) \erm^{2 i d n_s \omega} \nonumber \\
& & - (n_s+1)(n_s+1+\sigma_{xx}) \, , 
\nonumber \\
A_y &=& \sigma_{xy} \left[ n_s + 1 + (n_s-1) \erm^{2 i d n_s \omega} \right] . \nonumber
\eea
Note that the Hall conductivity $\sigma_{xy}$ is different from zero in presence of a non-vanishing magnetic field.

A first general result comes when considering two phenomenologically distinct explicit cases, that is suspended graphene samples without gate voltage, characterized by a small chemical potential, or epitaxial graphene, where a large Fermi energy shift is present due to the interaction with the atoms of the substrate. For the first case ($n_s=1$), the transmission coefficient reduces to 
\be \label{tsusp}
T^{\rm \, suspended} = 4 \, \frac{|\sigma_{xx} + 2 |^2 + |\sigma_{xy}|^2}{| (\sigma_{xx} + 2)^2 + \sigma_{xy}^2|^2} \, ,
\ee
while for epitaxial graphene ($n_s \neq 1$), when the intensity is averaged over Fabry-Perot oscillations \footnote{In standard cases, the rapid oscillations with the frequency $\omega$ or with the substrate width $d$ are smeared out due to the low resolution of experiments or to other sources of incoherence.}, we have:
\be \label{tepi}
T^{\rm \, epitaxial} = 8 n_s^2 \left( \frac{1}{|a_+|^2 - |b_+|^2} + \frac{1}{|a_-|^2 - |b_-|^2} \right) ,
\ee
\bea
a_\pm &=& \left( 1 + n_s \right) \left( 1 + n_s + \sigma_{xx} \pm i \sigma_{xy} \right) , \nonumber \\
b_\pm &=& \left( 1 - n_s \right) \left( 1 - n_s + \sigma_{xx} \pm i \sigma_{xy} \right) . \nonumber
\eea

\noindent After simple algebra, it can be easily deduced that, under equal material conditions, surprisingly
\be \label{comp}
T_{n_s \neq 1} > T_{n_s = 1} \, ,
\ee
that is transmission of light through graphene in presence of a substrate is {\it always} greater than with no substrate. {Such a result, depending only on the value of $n_s$ (within the tight-binding approach here adopted), is entirely due to the fact that, when a substrate is present, additional light interaction with it makes the phenomenon non-linear. In the Dirac model of quasiparticles, additional light reflected from the substrate and impinging back on graphene enhances transmission through graphene itself while diminishing reflection, the role played by multiple reflections being likely similar to that discussed in Ref. \cite{multi} in a different context. The net result is the effective opening of new transmission channels at the expenses of existing reflection ones \footnote{{Note that we are also assuming that no absorption takes place.}}: the reverse of relation (\ref{comp}) is never satisfied within our approximations. This effect is genuinely due to the action of the applied magnetic field. Indeed, the relation in (\ref{comp}) becomes an equality, i.e. $T_{n_s \neq 1} = T_{n_s = 1}$, only when $b_+=b_-=0$, that is, for $n_s \neq 1$, when $\sigma_{xy}=0$ and $\sigma_{xx}= n_s-1$, the Hall conductivity $\sigma_{xy}$ vanishing for zero applied field.}

An even lesser obvious result, again obtained without entering the details of ac conductivities, can be analytically deduced for the minimum transmission $T_{\rm min}$. Indeed, just by using the well-known triangle inequality, $|A_{x}^2| + |A_{y}^2| \geq |A_x^2 + A_y^2|$, we easily find that 
\be  \label{tmin}
T \geq \frac{16 n_s^2}{|A_{x}|^2 + |A_{y}|^2} \, ,
\ee
i.e. a {\it non-vanishing} minimum transmission $T_{\rm min} > 0$ is always present in graphene, in any condition. 

Further general results come when calculations are performed at first order in the fine structure constant $\alpha$ (by expanding the complex conductivities in terms of $\alpha$ and keeping just linear terms); in such a case we have:
\be \label{fo}
T \simeq \frac{2 n_s^2}{1 + n_s^2} \left[ 1 - \frac{3 + n_s^2}{2(1+n_s^2)} \, {\rm Re} \, \sigma_{xx} \right] .
\ee
It is easy to prove that, at first order in $\alpha$, 
\be \label{tmin2}
T \simeq T_{\rm min}
\ee
for both suspended and epitaxial graphene, so that when higher order corrections come into play they always tend to {\it increase} light transmission.

All such general results, coming just from applying the Dirac model, are obviously useful when designing given experiments, but more detailed predictions are required for direct physical applications. In order to achieve this, we need the explicit expressions for the ac conductivities; at first order in $\alpha$ we have \cite{Fialko2012}:
\bea
\sigma_{xx} &=& - 4 i \, \alpha \, v_F^2 \, e B \, (\omega + 2 i \Gamma) \sum_{n=0}^{n_{max}} \left[ \left( 1 - \frac{\Delta^2}{M_n M_{n+1}} \right) \right. \nonumber \\
& & \!\! \!\! \!\! \!\! \!\! \cdot \, \frac{n_F(M_n) - n_F(M_{n+1}) + n_F(-M_{n+1}) - n_F(-M_n)}{[(M_{n+1}-M_n)^2 - (\omega + 2 i \Gamma)^2 ] (M_{n+1} - M_n)} \nonumber \\ 
& &  \left.+ \left\{ M_n \rightarrow - M_n ; \,\, M_{n+1} \rightarrow M_{n+1} \right\} \right] , \nonumber
\\ & & \label{sigma} \\
\sigma_{xy} &=& - 4 \, \alpha \, v_F^2 \, e B  \sum_{n=0}^{n_{max}}  \left[ n_F(M_n) -n_F(M_{n+1}) \right.
\nonumber \\  
& & \!\! \!\! \!\! \!\! \!\! \left. - n_F(-M_{n+1}) + n_F(-M_n) \right] \left[ \left( 1 - \frac{\Delta^2}{M_n M_{n+1}} \right) \right. \nonumber \\
& & \!\! \!\! \!\! \!\! \!\! \cdot \, \frac{1}{(M_{n+1}-M_n)^2 - (\omega + 2 i \Gamma)^2} \nonumber \\
& & \left. + \left\{ M_n \rightarrow - M_n , \,\, M_{n+1} \rightarrow M_{n+1} \right\} \right] , \nonumber
\eea
where $n_F(x) = [\erm^{(x-\mu)/t} +1]^{-1}$ is the Fermi function at temperature $t$, $M_n = \sqrt{2 n v_F^2 e B + \Delta^2}$ are the Landau levels for an applied magnetic field $B$ (perpendicular to the graphene sheet), and $n_0$ is the integer part of $(\mu^2- \Delta^2)/2 v_F^2 e B$. %, $n_{max}=n_0+30$. 
For fixed temperature and given magnetic field, the expressions above depend on three parameters (in addition to the substrate refractive index), namely the chemical potential $\mu$, the mass gap $\Delta$ and a phenomenological (constant, positive) parameter $\Gamma$ describing the presence of impurities in realistic graphene samples by means of the substitution $\omega \rightarrow \omega + i \Gamma {\rm sgn} \, \omega$. For the sake of illustration, below we choose typical numerical values for these parameters as coming from the analysis performed in \cite{Fialko2012} of the experiment reported in Ref. \cite{Grassee2011} (graphene sample on a SiC substrate) for which the validity of the Dirac model has been established.

\begin{figure}
\begin{center}
\bt{ll}
\vspace{-2truemm}
\includegraphics[width=4.2cm]{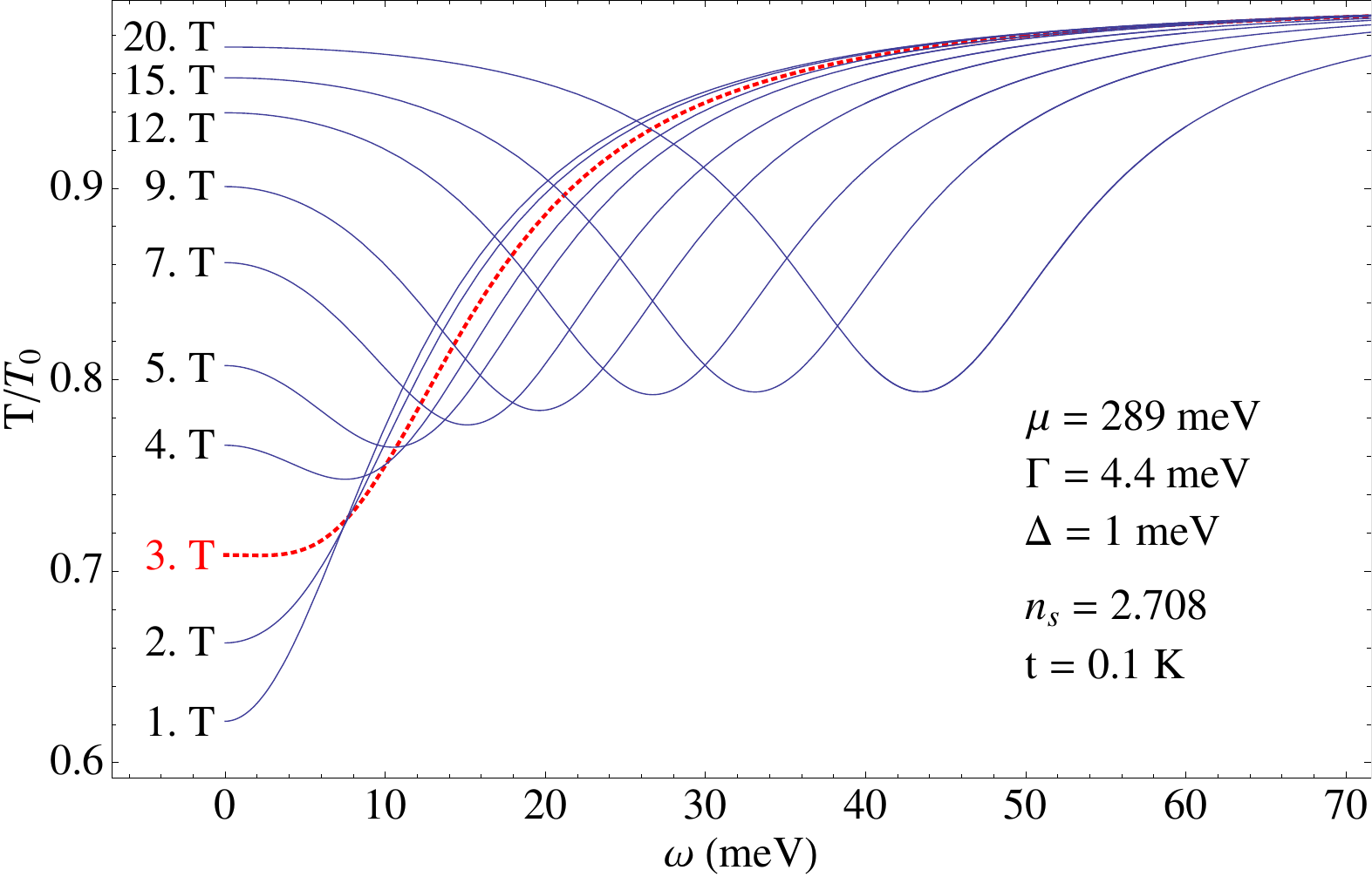}  &  \includegraphics[width=4.2cm]{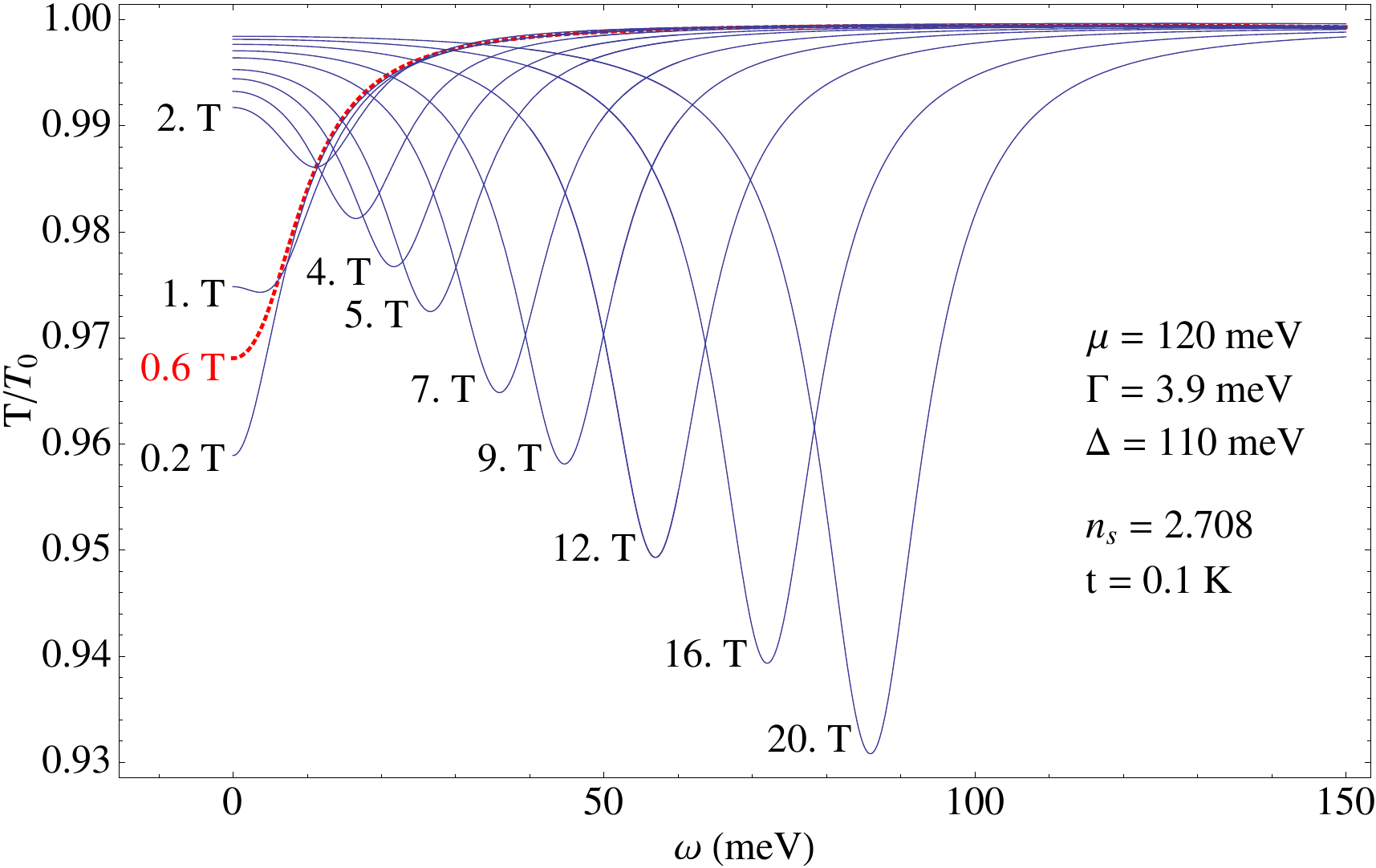} \\  
a) & b) \\ & \\
\vspace{-2truemm}
\includegraphics[width=4.2cm]{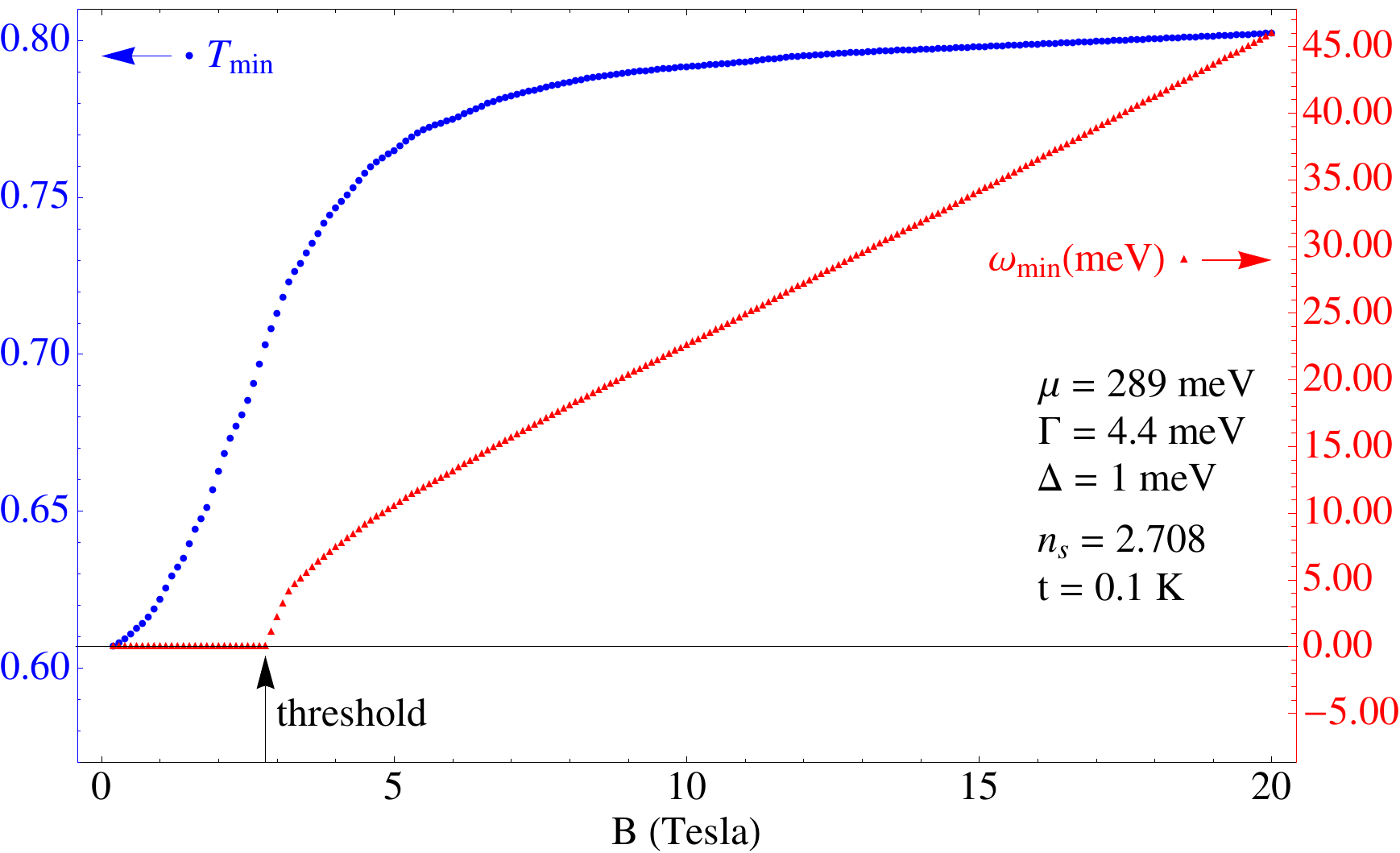}  & \includegraphics[width=4.2cm]{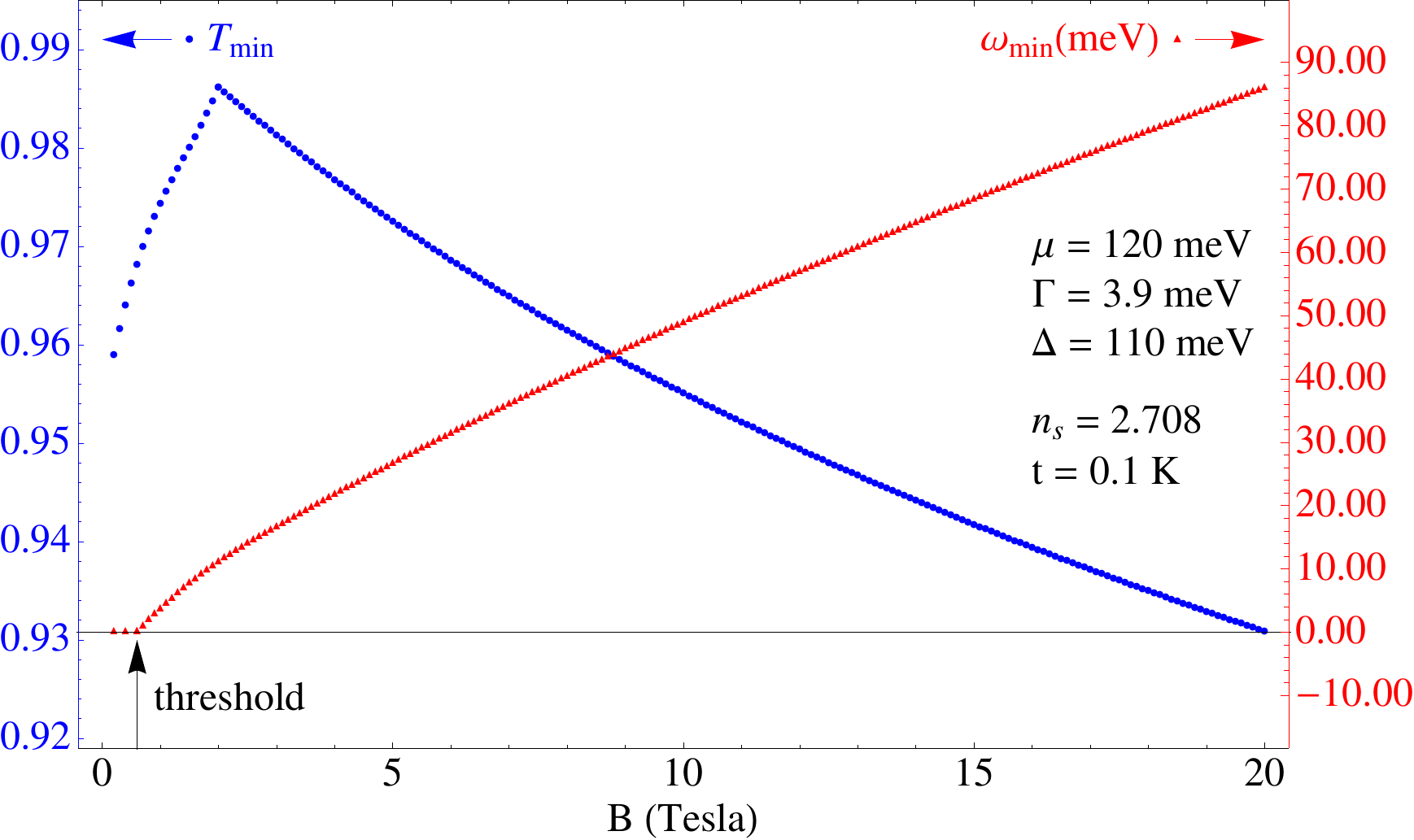} \\
c) & d)
\et
\caption{a), b) Transmission coefficient of light in a magnetized epitaxial graphene sample, as a function of light frequency $\omega$ for different values of the applied magnetic field. c), d) Minimum value of the transmission (left scale), and corresponding value of the light frequency (right scale), as a function of the applied magnetic field. A threshold effect evidently appears for a non-trivial realization of the minimum transmission.}
\label{fig1}
\end{center}
\end{figure}

\begin{figure}
\begin{center}
\bt{ll}
\vspace{-2truemm}
\includegraphics[width=4.2cm]{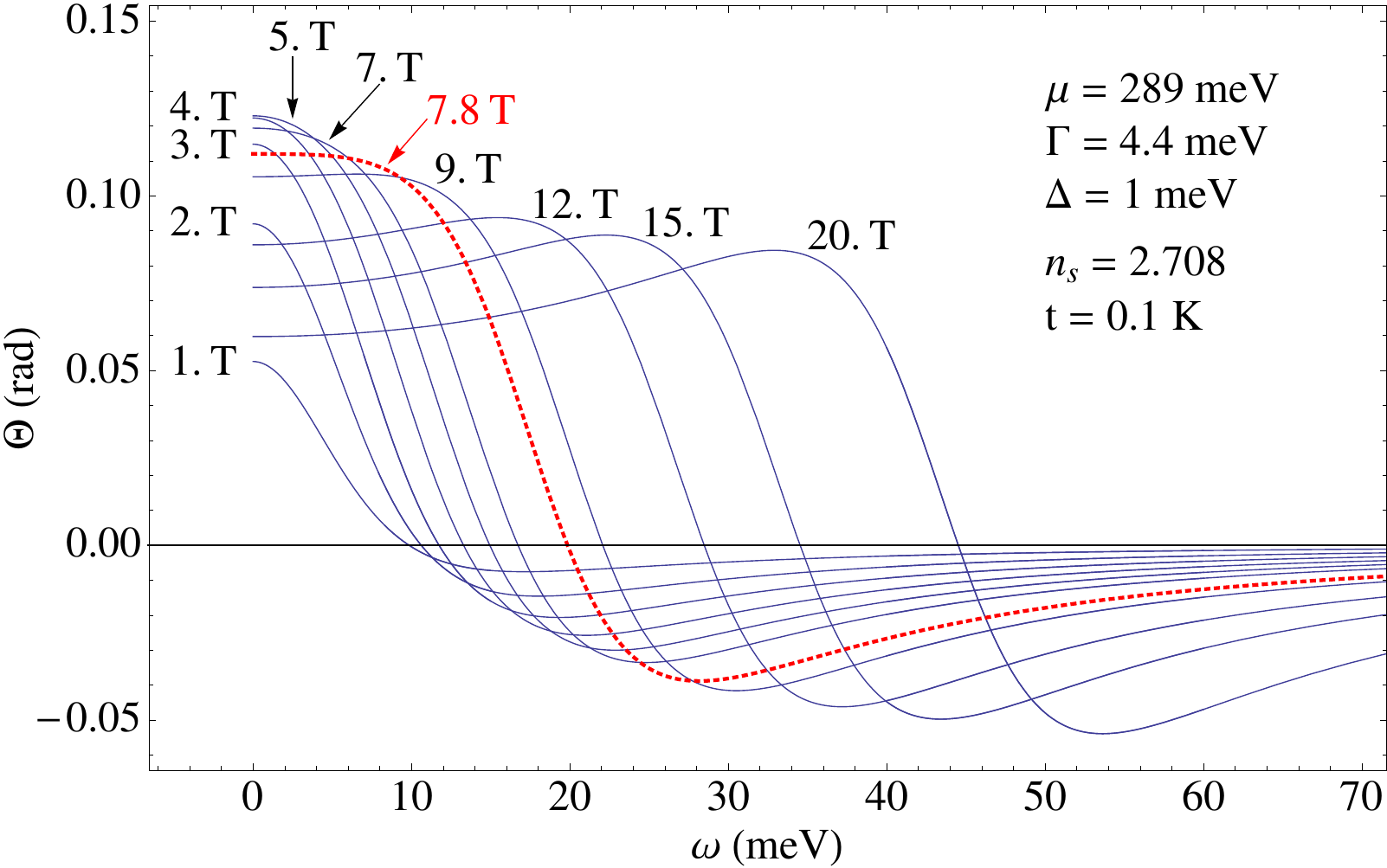}  & \includegraphics[width=4.2cm]{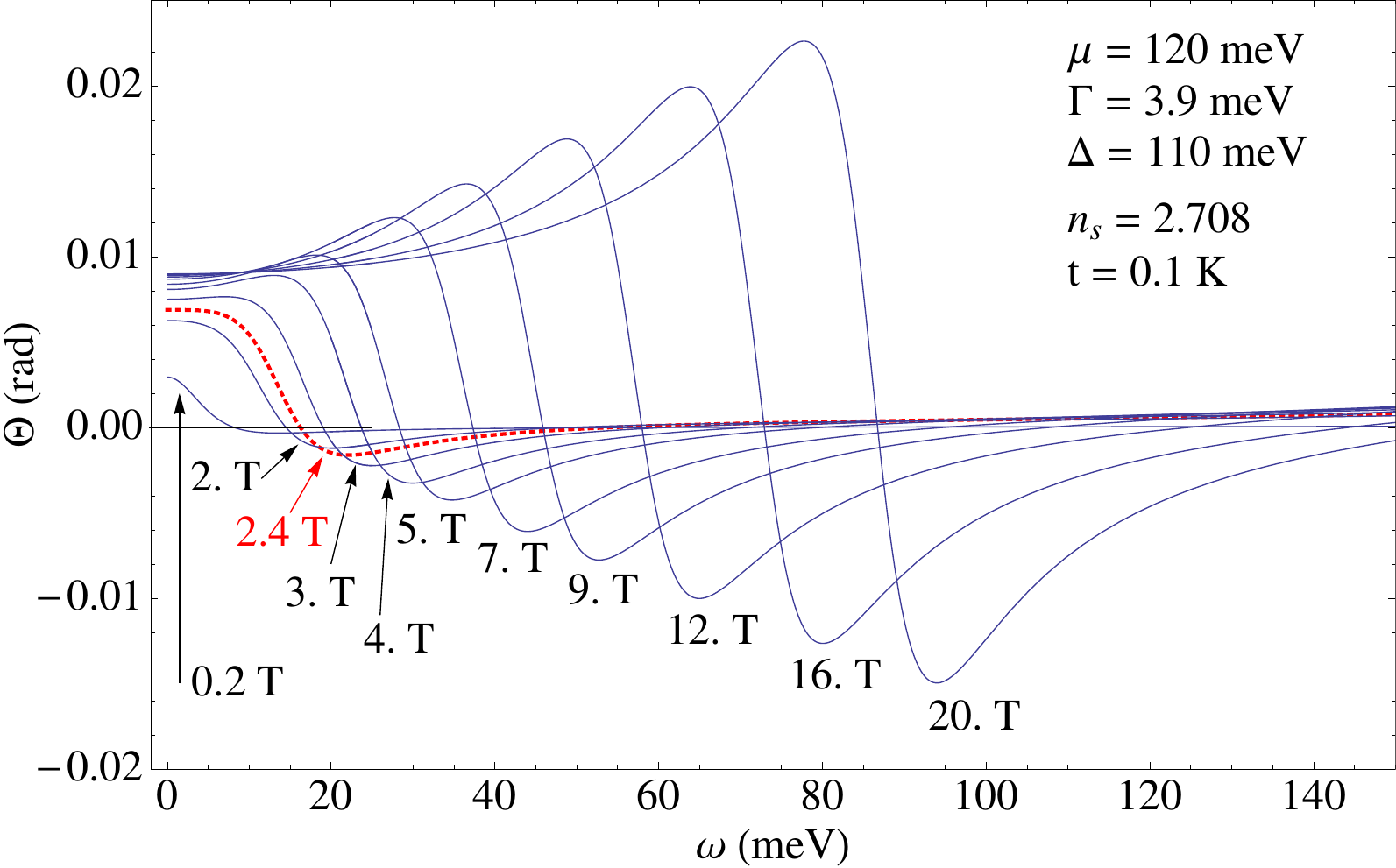} \\ 
a) & b) \\ & \\
\vspace{-2truemm}
\includegraphics[width=4.2cm]{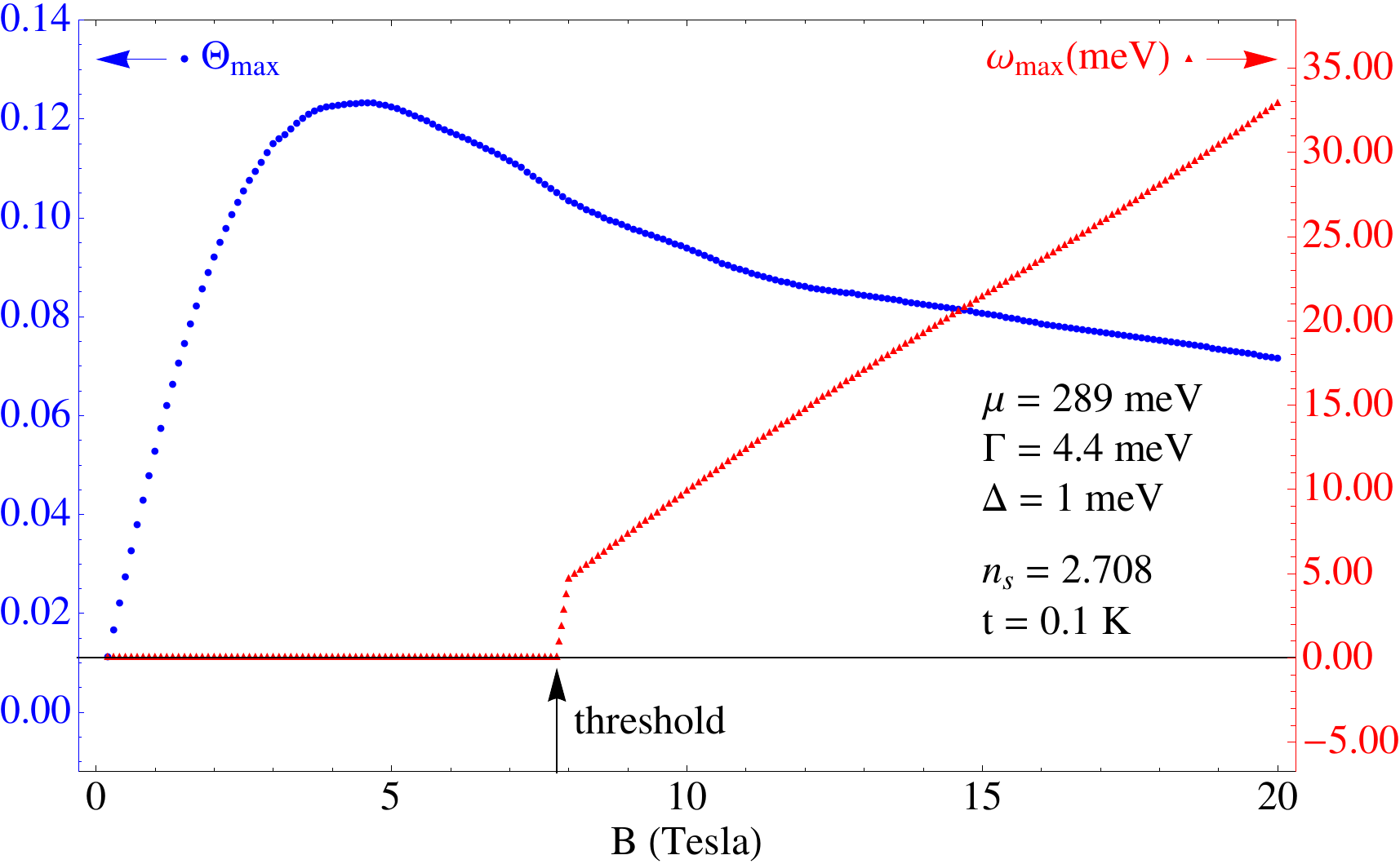}  & \includegraphics[width=4.2cm]{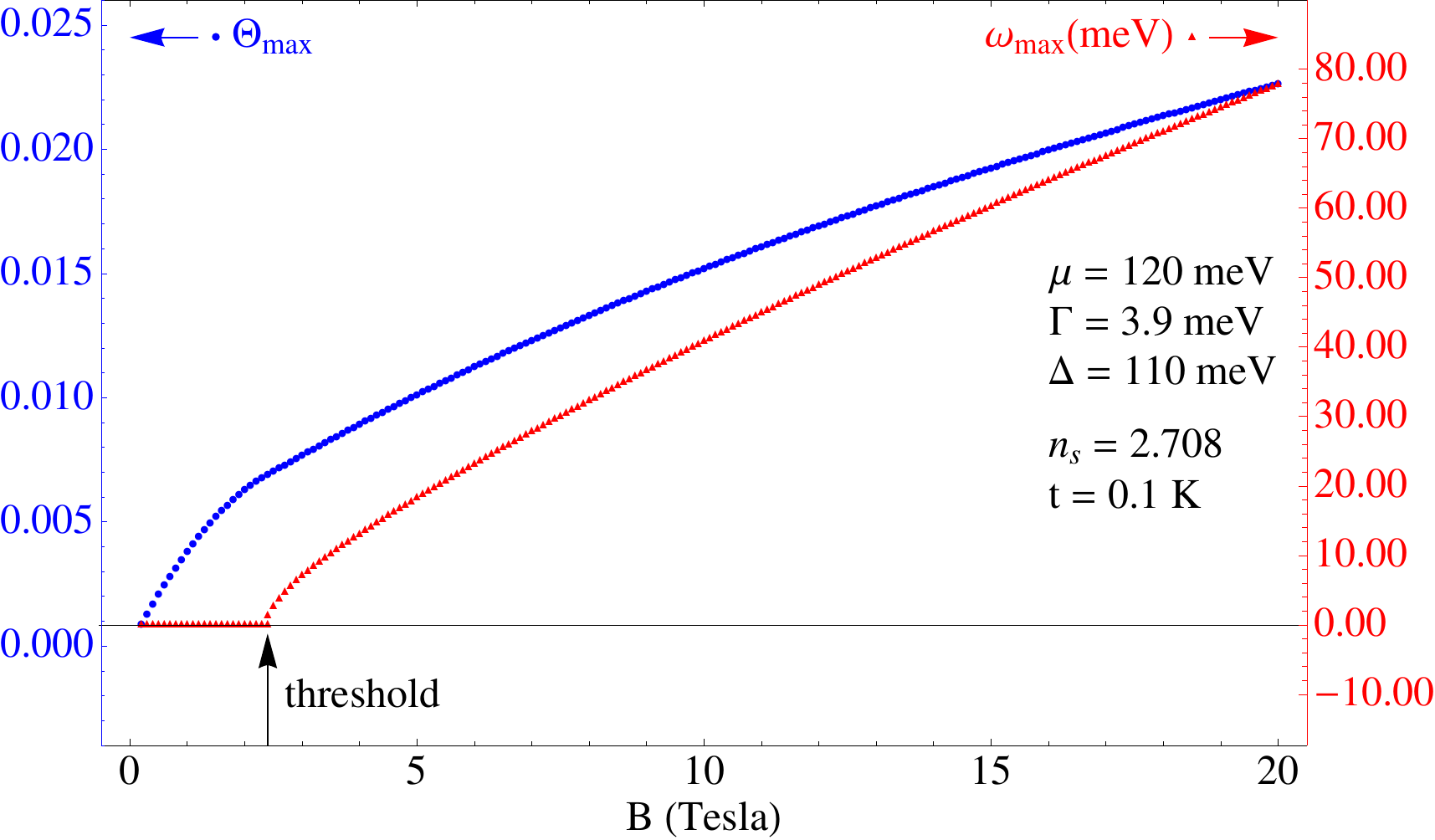} \\  
c) & d) \\ & \\
\vspace{-2truemm}
\includegraphics[width=4.2cm]{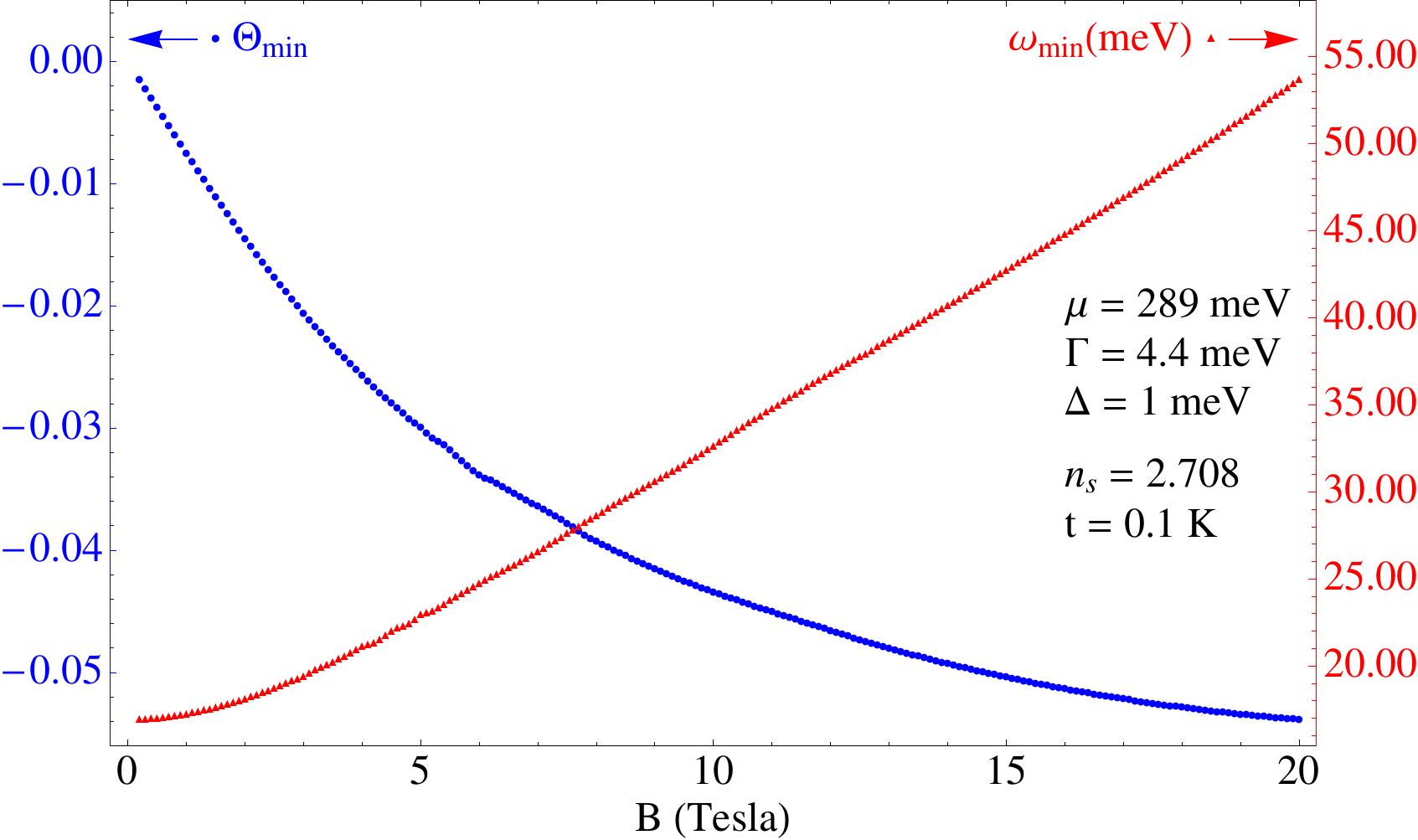}  & \includegraphics[width=4.2cm]{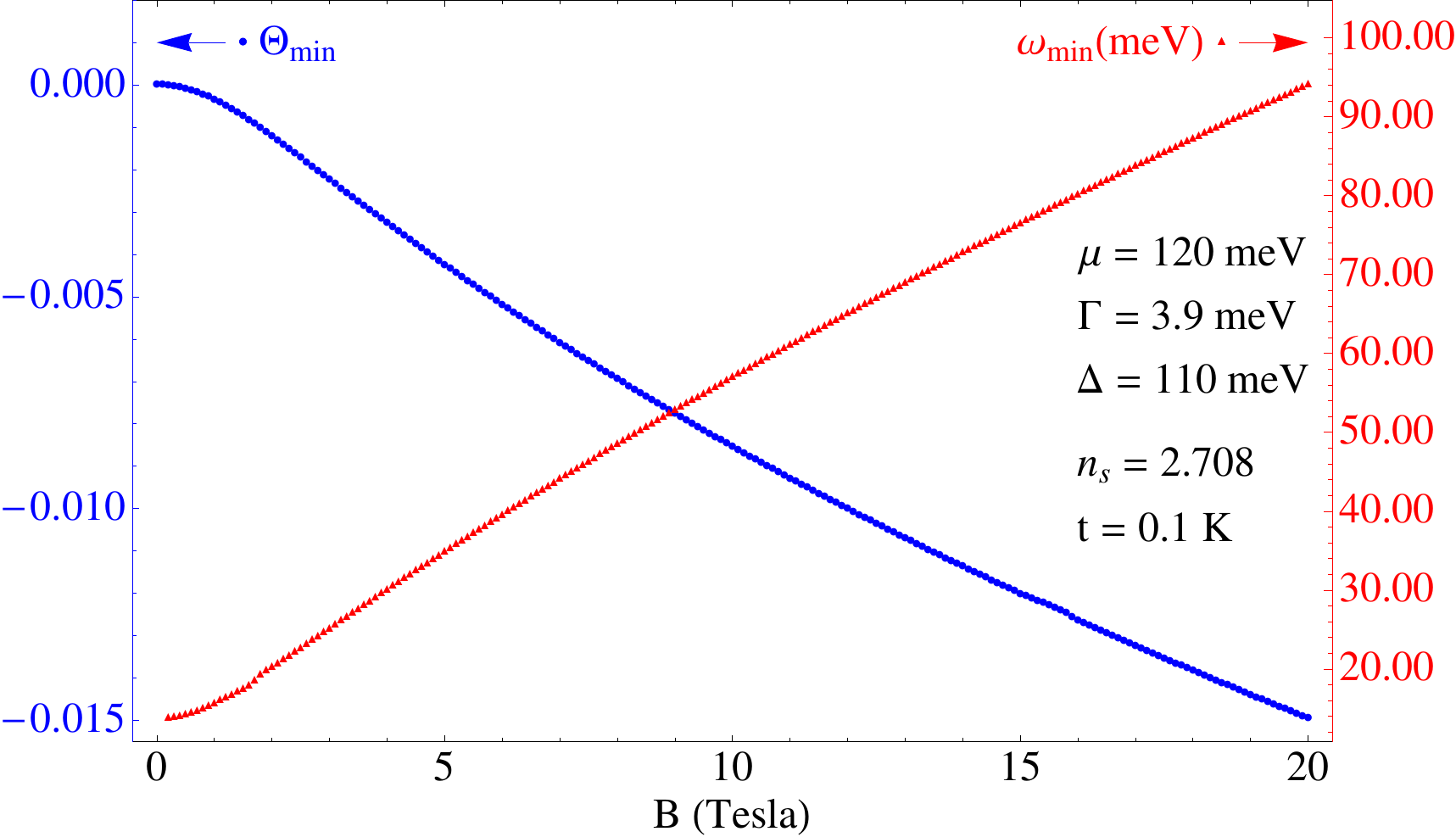} \\
e) & f)
\et
\caption{a), b) Faraday rotation angle in a magnetized epitaxial graphene sample, as a function of light frequency $\omega$ for different values of the applied magnetic field. c), d) Maximum and e), f) minimum value of the polarization angle (left scale), and corresponding values of the light frequency (right scale), as a function of the applied magnetic field. A threshold effect manifests for a non-trivial realization of maximum Faraday rotation.}
\label{fig2}
\end{center}
\end{figure}

The transmission coefficient (normalized to the value of the bare substrate, $T_0=2n_s/(1+n_s^2)$) as a function of light frequency, for different values of the applied magnetic field, is plotted in Fig.s\,\ref{fig1}a,\ref{fig1}b for two different regimes. The minimum of the transmission always increases for increasing magnetic field (in the frequency range considered) in the large chemical potential case, while decreases for not small magnetic fields in the moderate chemical potential regime (after the reaching of a relative maximum). The unexpected result is the presence of a {\it threshold effect}: a non-trivial minimum transmission corresponding to a non-zero frequency manifests only for an applied magnetic field {\it greater} than a critical value (see Fig.\,\ref{fig1}c,\ref{fig1}d). Such a threshold field is higher (and then more easily detectable) for higher chemical potentials.

A similar effect is present in the Faraday rotation angle of light polarization through graphene, which can be evaluated \cite{Fialko2012} from
\be \label{theta}
\theta = - \frac{1}{2} \arctan \, \frac{8 \, {\rm Re} (\sigma_{xx} \sigma_{xy}^\ast) + 4 (n_s^2 + 3) \, {\rm Re} \, \sigma_{xy}}{|2 \sigma_{xx} + n_s^2 + 3|^2 - 4 |\sigma_{xy}|^2 - (n_s^2 -1))^2 } \, ,
\ee
with the ac conductivities reported above. This rotation angle of light polarization is plotted in Fig.s\,\ref{fig2}a,\ref{fig2}b for different values of the material parameters as a function of the light frequency, showing the ``giant Faraday effect'' observed in Ref. \cite{Grassee2011} (and theoretically found in \cite{Fialko2012}). In the case depicted in Fig.\,\ref{fig2}a, maximum rotation increases with the magnetic field for low fields, up to a maximum around $4\div5$ T, but this is accomplished only for the trivial case $\omega =0$. Instead a threshold effect manifests for a critical magnetic field of $7.8$ T, after which maximum rotation always decreases, being reached for increasing values of light frequency (Fig.\,\ref{fig2}c). Quite a similar effect displays also for moderate chemical potentials (Fig.\,\ref{fig2}d), although the threshold field is lower ($2.4$ T) and the maximum rotation always increases with the applied magnetic field, for increasing values of the light frequency. The polarization angle curves also present (negative) minima which, for any value of the chemical potential (see Fig.s\,\ref{fig2}e,\ref{fig2}f), always decrease with increasing magnetic field, while increasing the light frequency, no threshold effect being present.

\begin{figure}
\begin{center}
\bt{ll}
\vspace{-2truemm}
\includegraphics[width=4.2cm]{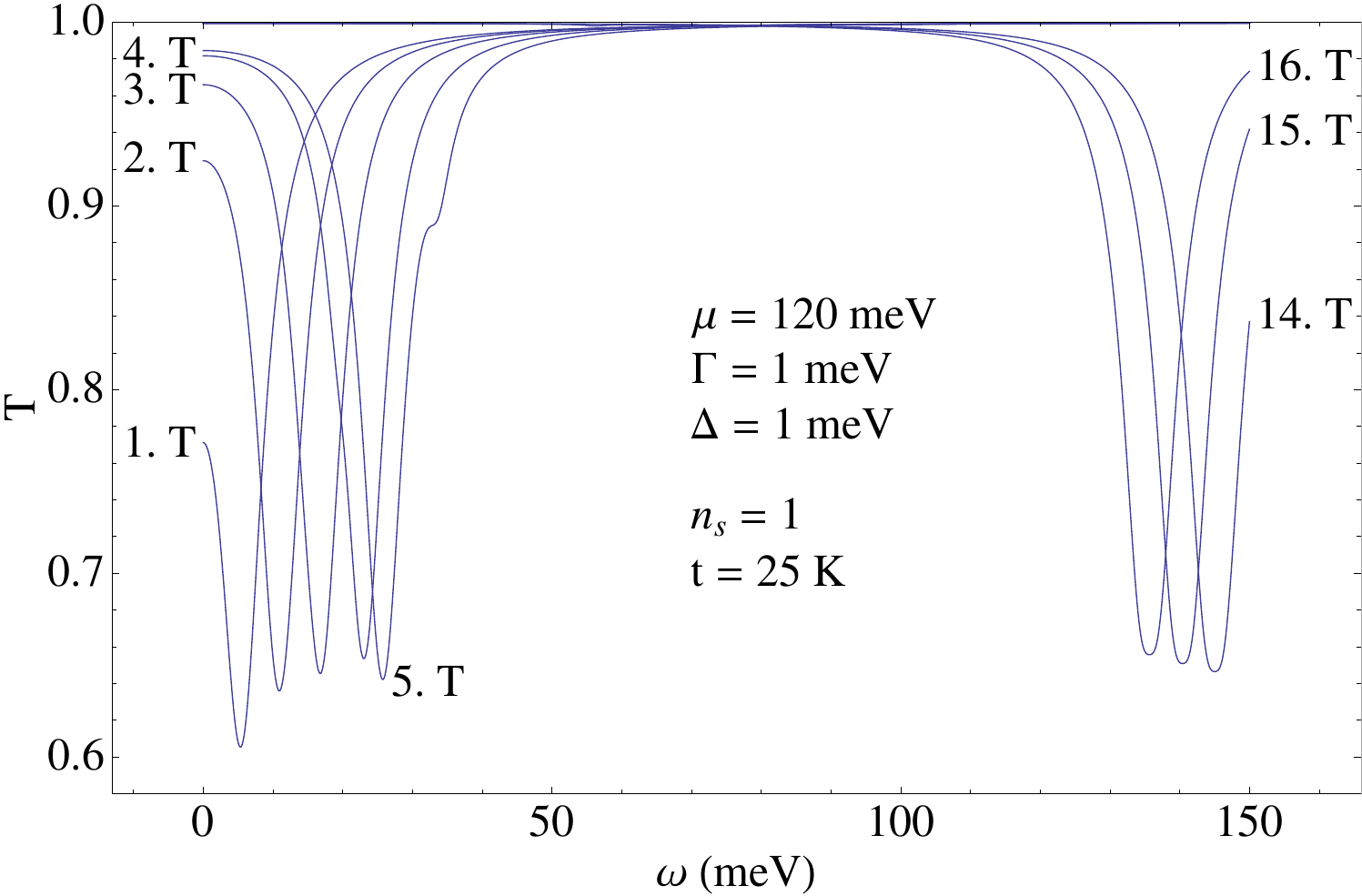}  & \includegraphics[width=4.2cm]{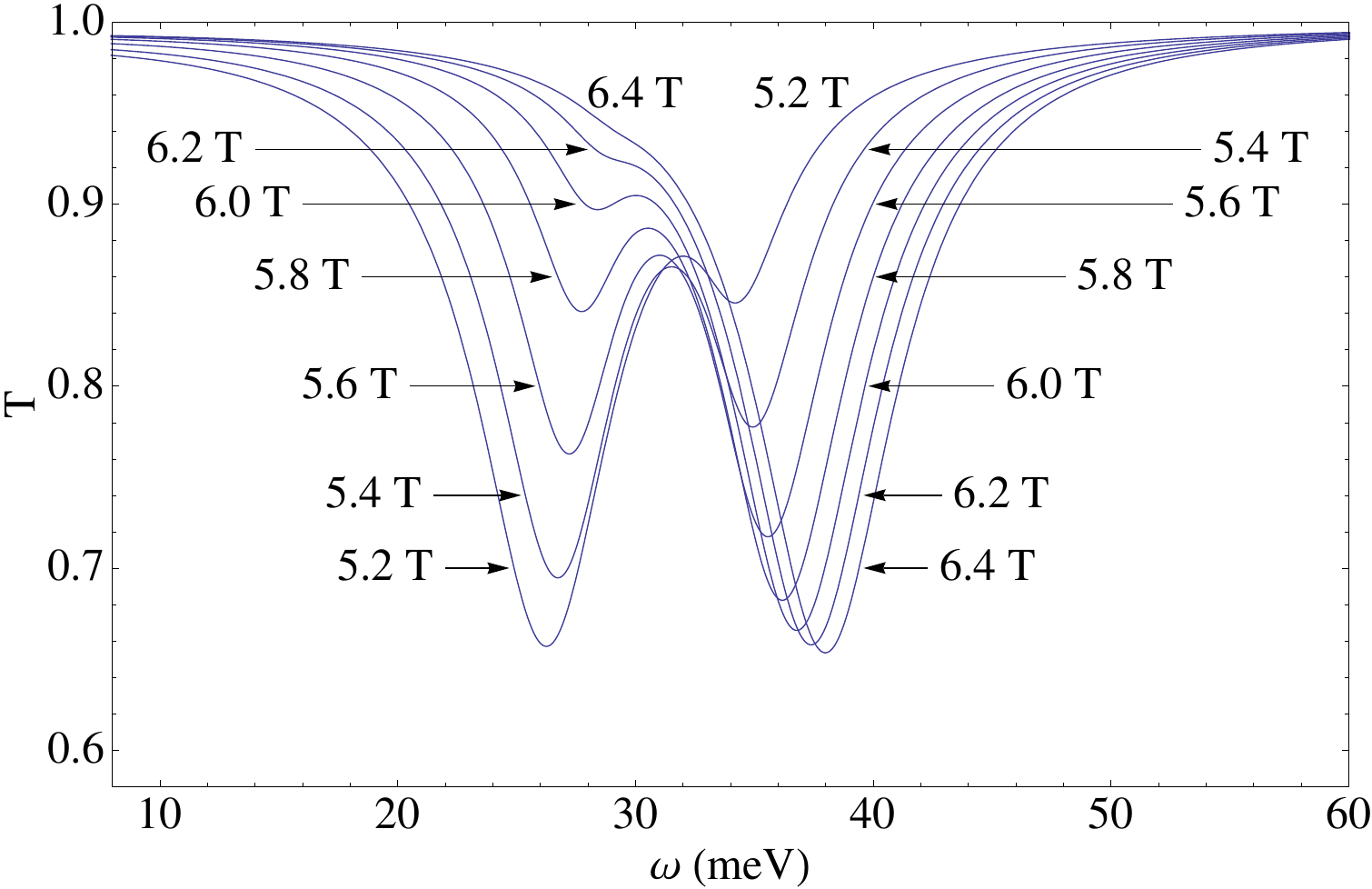} \\ 
a) & b) \\ & \\
\vspace{-2truemm}
\includegraphics[width=4.2cm]{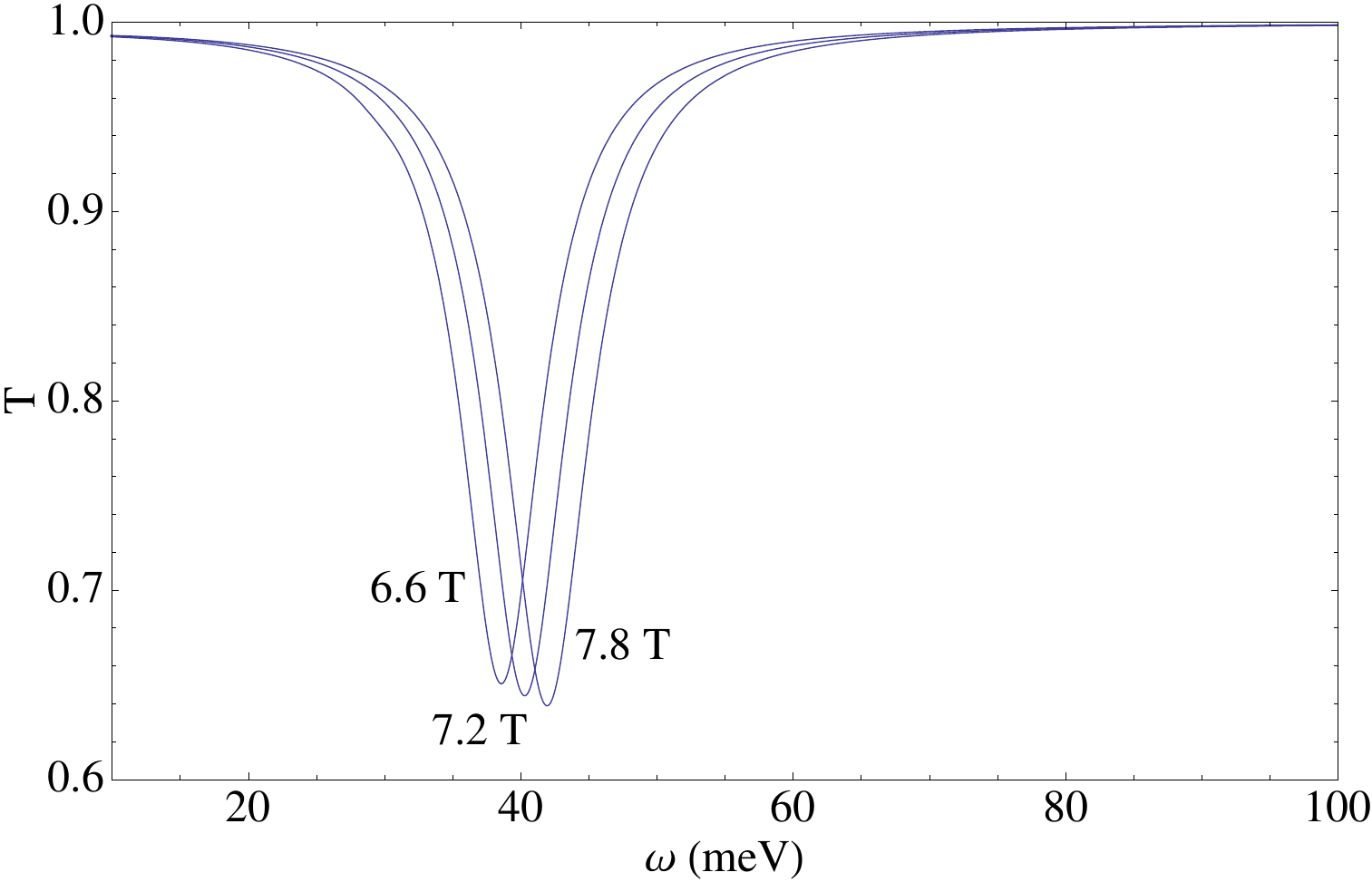}  & \includegraphics[width=4.2cm]{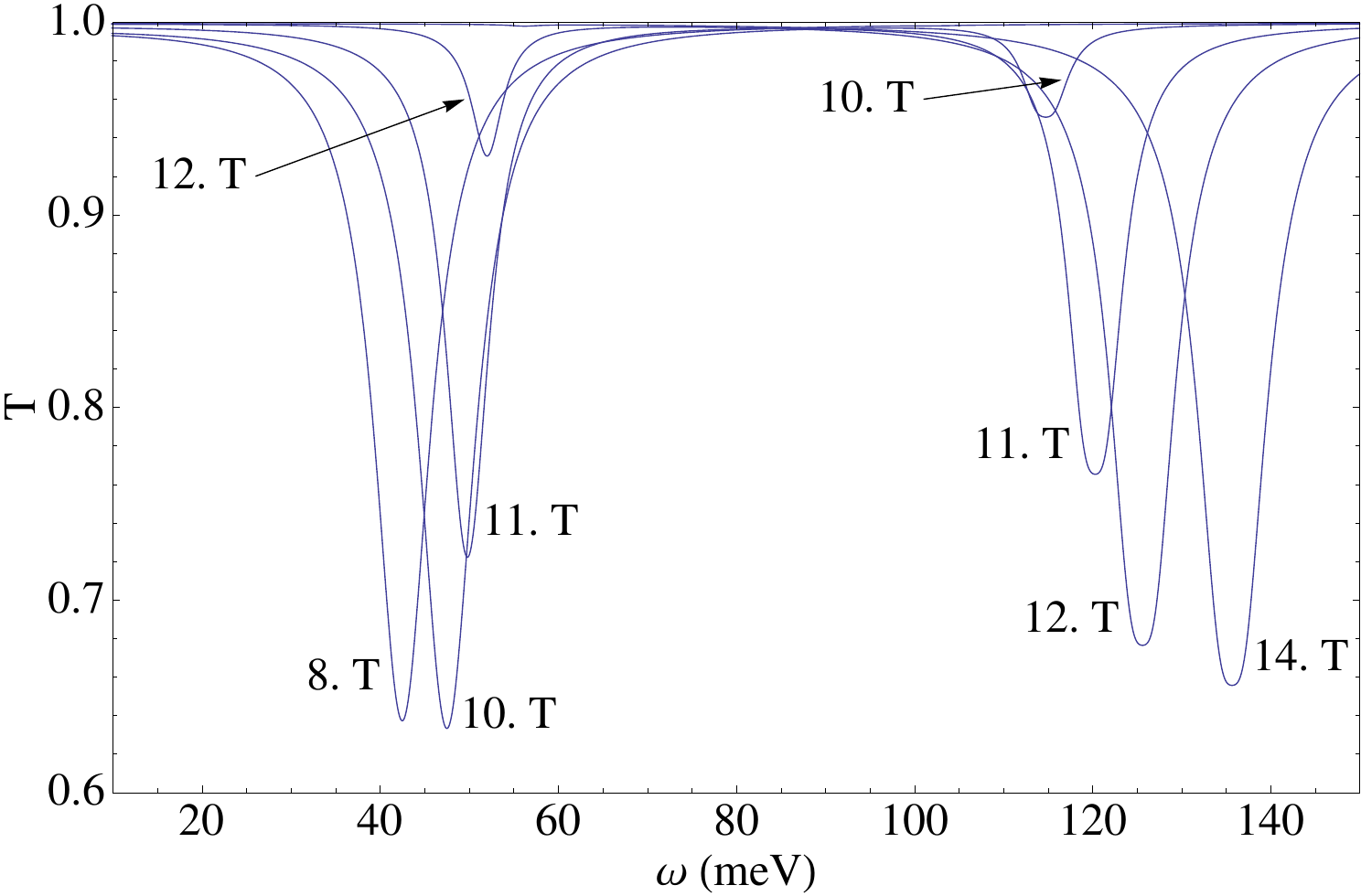} \\
c) & d)
\et
\caption{Transmission coefficient of light in a magnetized suspended graphene sample, as a function of light frequency $\omega$ for different values of the applied magnetic field. Note that the general condition in (\ref{comp}) is apparently fulfilled. Multiple minima appear for given values of the field, along with a ``shift effect'' towards higher values of the frequency.}
\label{fig3}
\end{center}
\end{figure}

The situation gets even more interesting in suspended graphene or, as studied experimentally in  Ref. \cite{Neugebauer}, when a decoupled graphene layer from the substrate material is considered. Indeed, more than one transmission minimum enters in the non-trivial frequency region of interest when the applied magnetic field increases, lower frequency minima giving way to higher frequency ones up to completely disappear, as it can be well noted in Fig.s\,\ref{fig5}. This leads to an effective {\it shift} of frequency bands where the sample gets more or less absorptive with a suitable tuning of the magnetic field. As an illustrative example, with typical values of the parameters (see Fig.\,\ref{fig5}), the first transition occurs with $B$ in the narrow range $5.4 \div 6.0$ T, producing a net frequency shift of $\Delta \omega \simeq 6$ meV (from about $\omega = 28$ meV to $\omega = 34$ meV), while the second broader transition with $B$ in the range $8.2 \div 14$ T leads to $\Delta \omega \simeq 47$ meV (with $\omega$ changing approximately from $56$ to $103$ meV).

\begin{figure}
\begin{center}
\includegraphics[width=8.5cm]{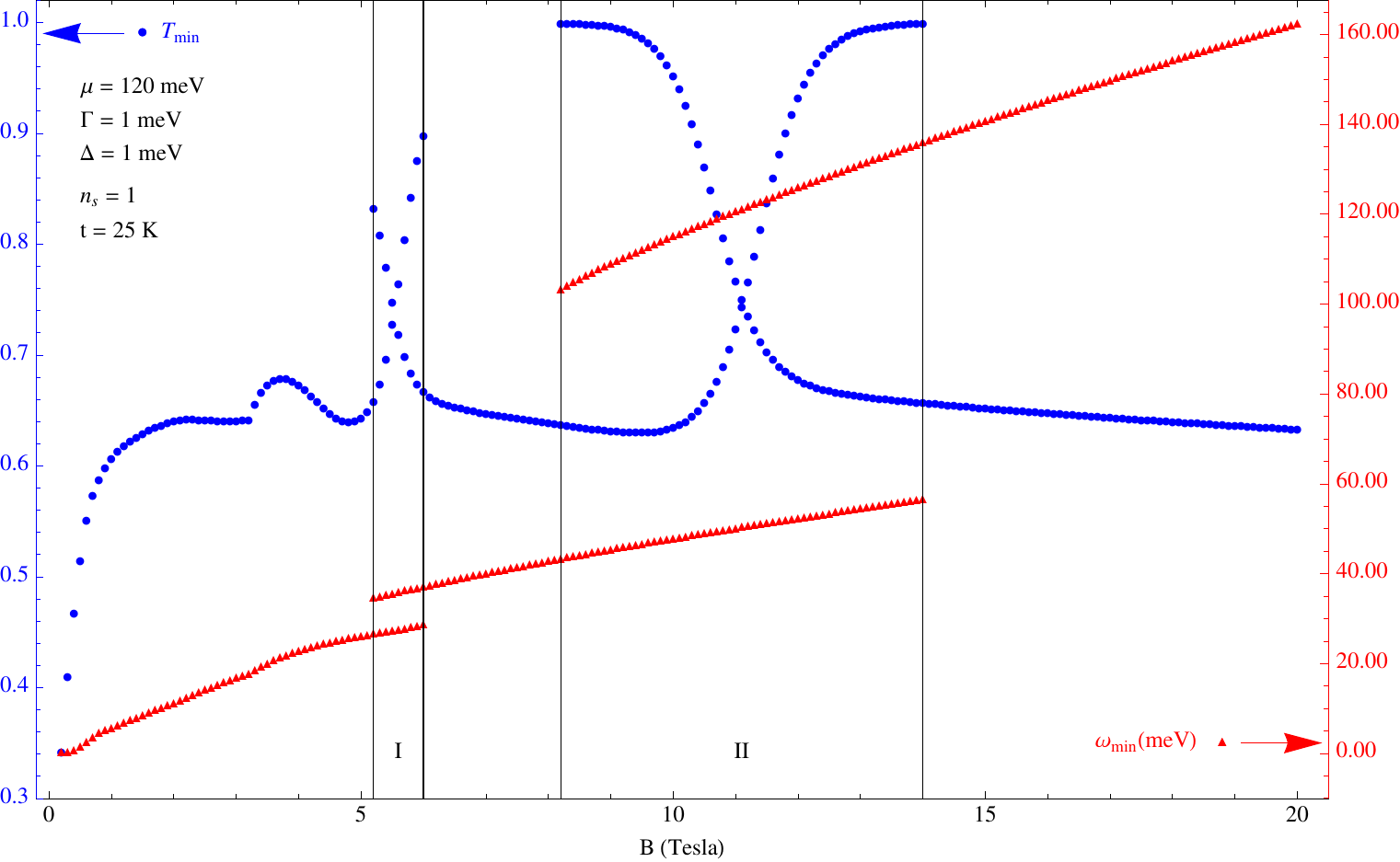}
\caption{Minimum transmission (left scale), and corresponding value of the light frequency (right scale), in a magnetized suspended graphene sample as a function of the applied magnetic field. The shift effect envisaged in Fig.\,\ref{fig3} takes place by tuning the field into two well defined transition regions (I and II).}
\label{fig4}
\end{center}
\end{figure}

A completely similar phenomenon takes place also for the Faraday rotation curves (see Fig.\,\ref{fig5}), the shift effect being here interrelated to the occurrence of minima and maxima. The two transition regions in the applied magnetic field (Fig.\,\ref{fig6}) are approximately the same as for minimum transmission: the first one is for $B$ in the range $5.0 \div 6.0$ T, producing a net frequency shift of $\Delta \omega = 8$ meV (from about $\omega=25$ meV to $\omega = 33$ meV), while the second one takes place for $B$ tuned in the region $8.8 \div 14.2$ T, leading to $\Delta \omega = 50$ meV (with $\omega$ changing approximately from $56$ to $106$ meV). The absence of a substrate in the graphene sample then allows to detect such an effect in the considered region of light frequency \footnote{Similar effects would, of course, occur also for epitaxial graphene samples but, due to the effective values of the material parameters (mainly, the chemical potential), they would produce only for very high values of the frequency of the electromagnetic radiation involved. In such cases, however, novel quantum field theory effects would be considered, which prevent the use of the expressions above for ac conductivities.}. 

The rich phenomenology reported above adds to the already large set of properties known for graphene, but it is quite remarkable that, just by tuning the value of an external magnetic field, the transmission and polarization features of light through the sample may change substantially. In particular, the possibility to have, preferentially in epitaxial graphene, an appreciable non trivial minimum transmission (whose presence is ensured by general theoretical arguments) in a desired frequency region controlled by an applied magnetic field is especially relevant for all those technological applications involving large THz and infrared detection, including graphene-based photovoltaics and ultrafast optoelectronic devices. On the other hand, avoiding such field regions results into more transparent samples (for moderate chemical potentials) that are particularly appropriate as protection layers for optical devices. For both kind of applications, the threshold effect for epitaxial graphene as well as the more intriguing shift effect for suspended graphene discussed above have to be taken in due account, in order not to undermine the efforts to reach the desired goal. Also, the plethora of intriguing features regarding the polarization properties of light absorbed by graphene is amenable of further experimental investigations for possible peculiar applications. Finally, the theoretical general predictions deduced above, irrespective of the particular specific model employed, reveal to be a very useful guidance in designing future phenomenological research about the unceasingly surprising graphene.

\begin{figure}
\begin{center}
\bt{ll}
\vspace{-2truemm}
\includegraphics[width=4.2cm]{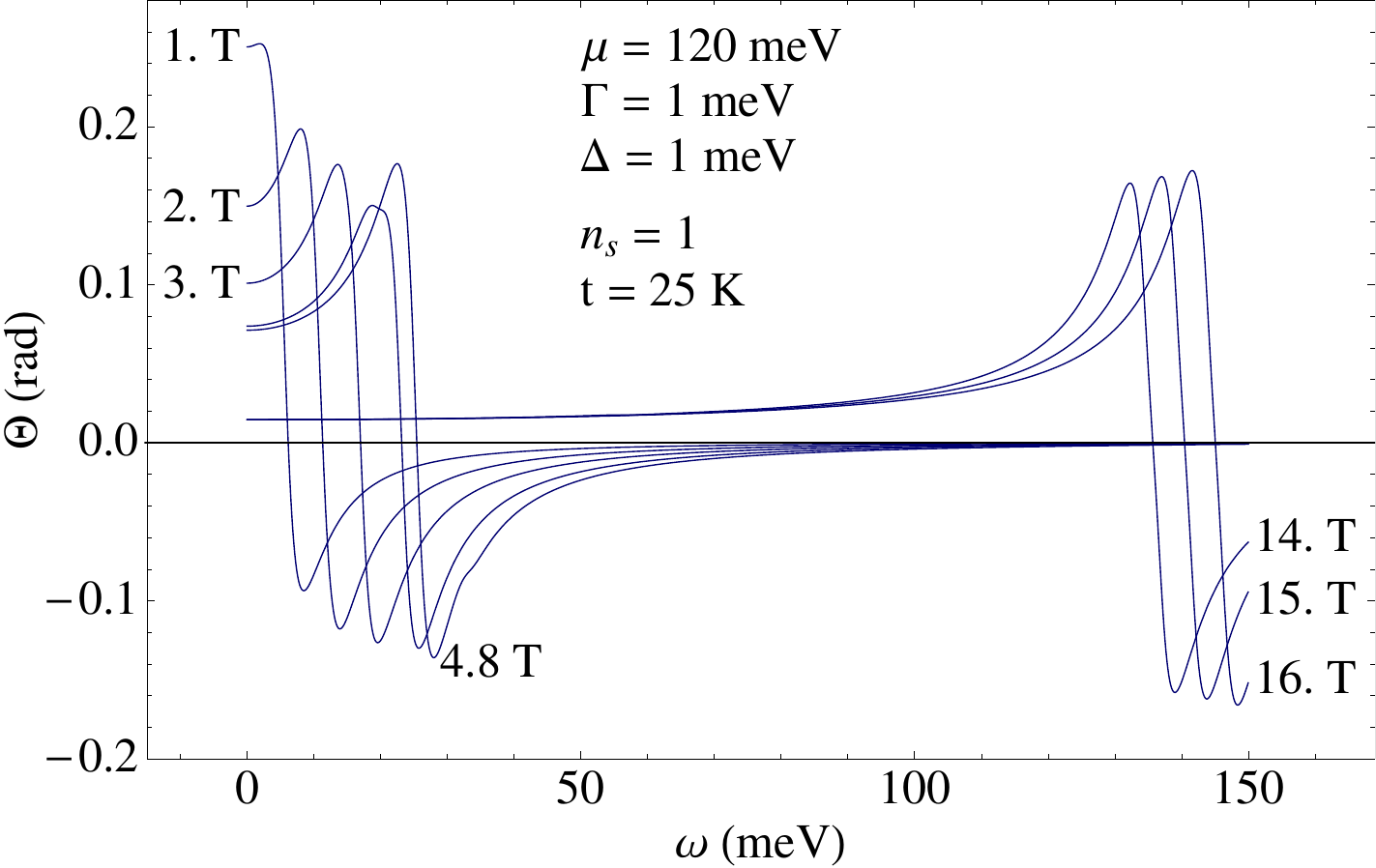}  & \includegraphics[width=4.2cm]{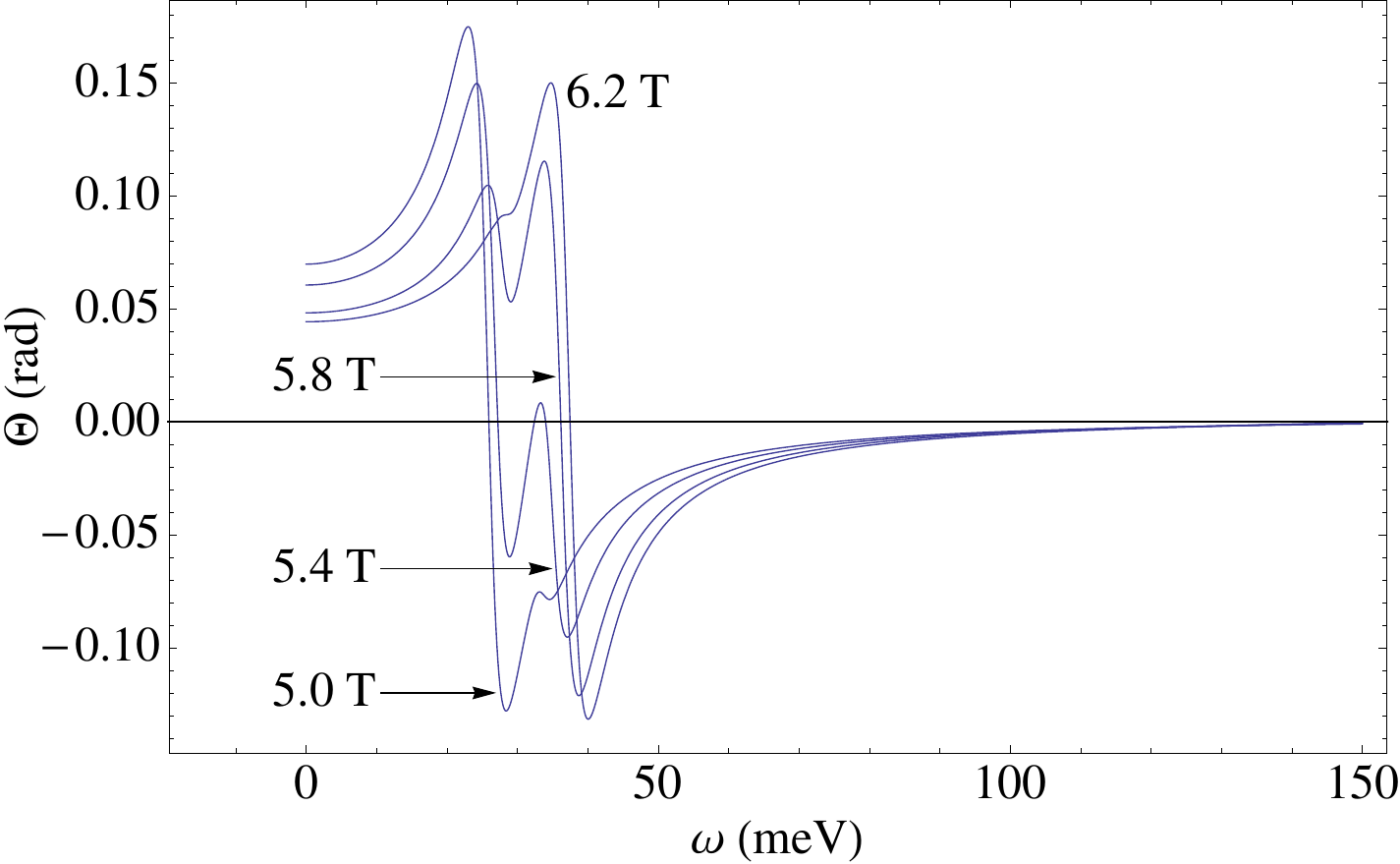} \\ 
a) & b) \\ & \\
\vspace{-2truemm}
\includegraphics[width=4.2cm]{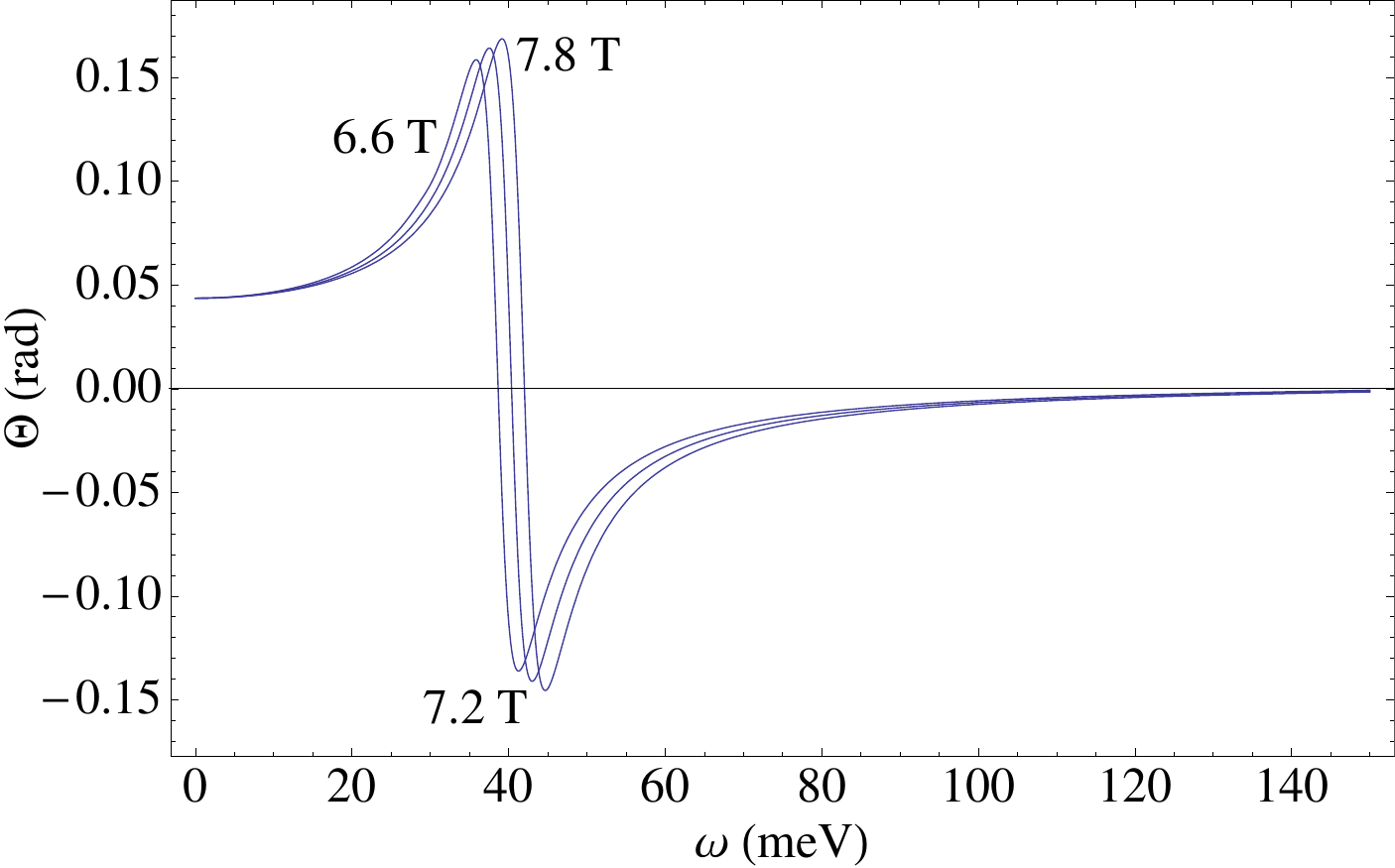}  & \includegraphics[width=4.2cm]{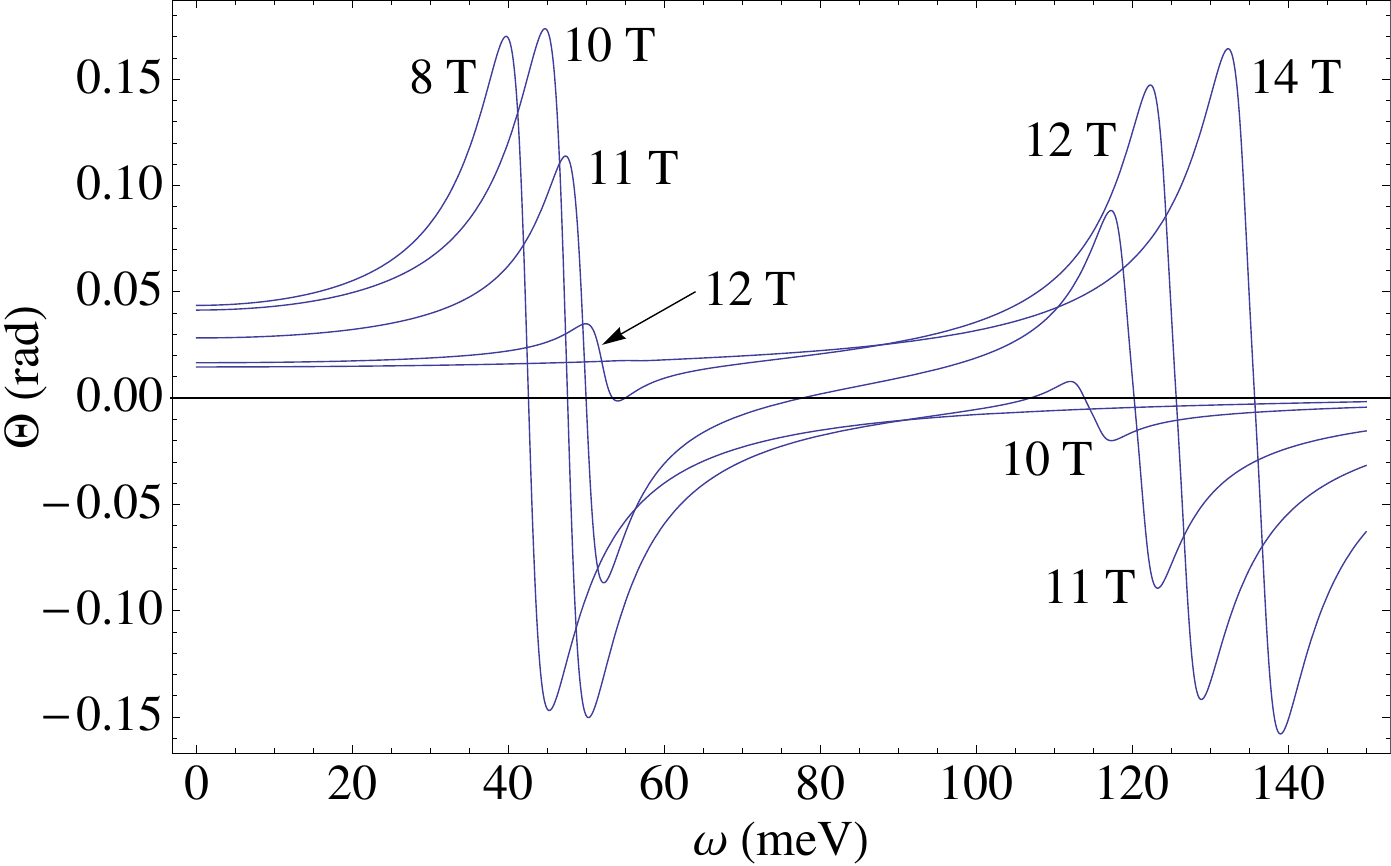} \\
c) & d) 
\et
\caption{Faraday rotation angle in a magnetized suspended graphene sample, as a function of light frequency $\omega$ for different values of the applied magnetic field. Multiple maxima and minima appear for given values of the field, along with a shift effect (for both maxima and mionima) towards higher values of the frequency.}
\label{fig5}
\end{center}
\end{figure}

\begin{figure}
\begin{center}
\bt{ll}
\vspace{-2truemm}
\includegraphics[width=4.2cm]{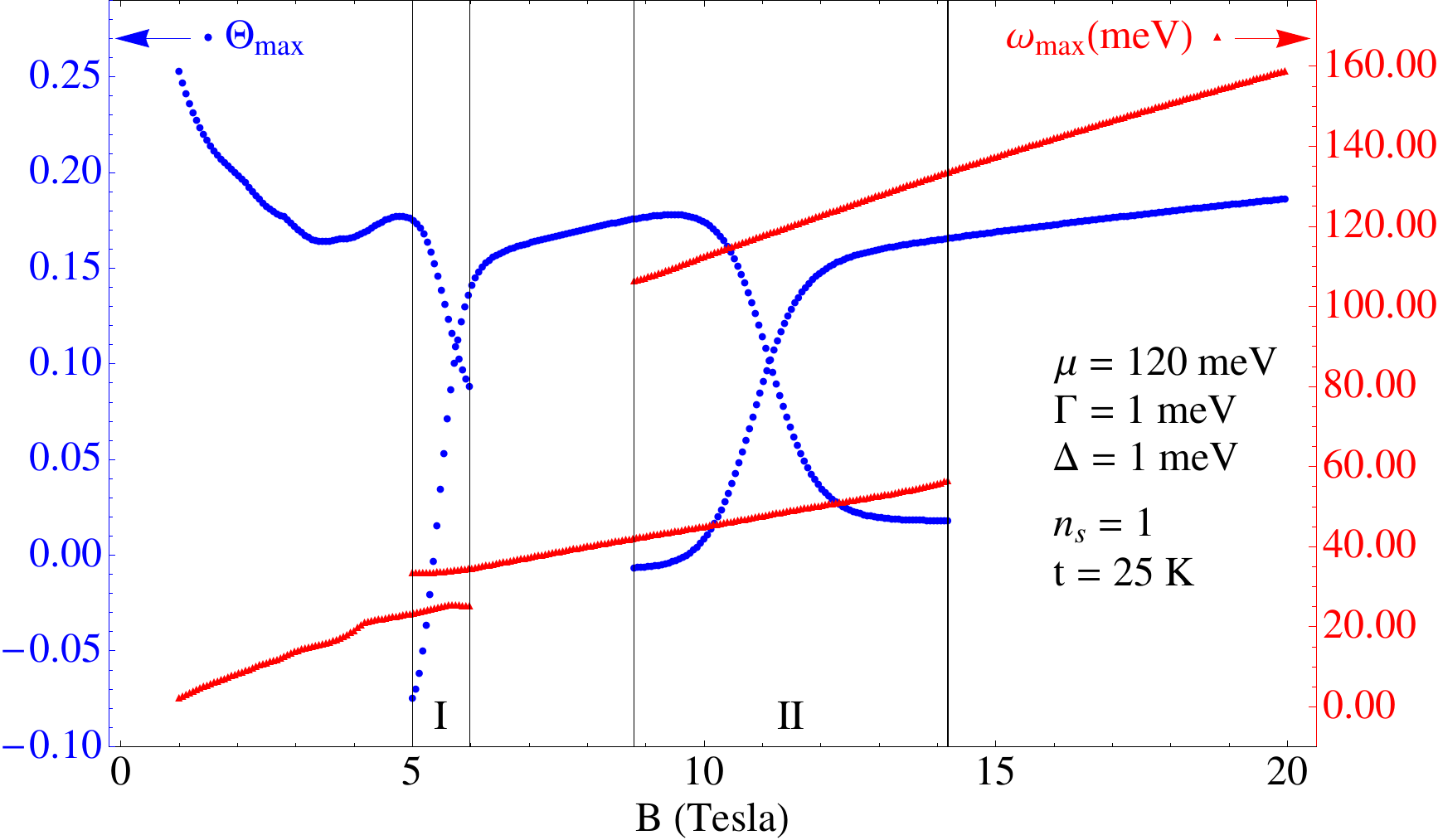}  & \includegraphics[width=4.2cm]{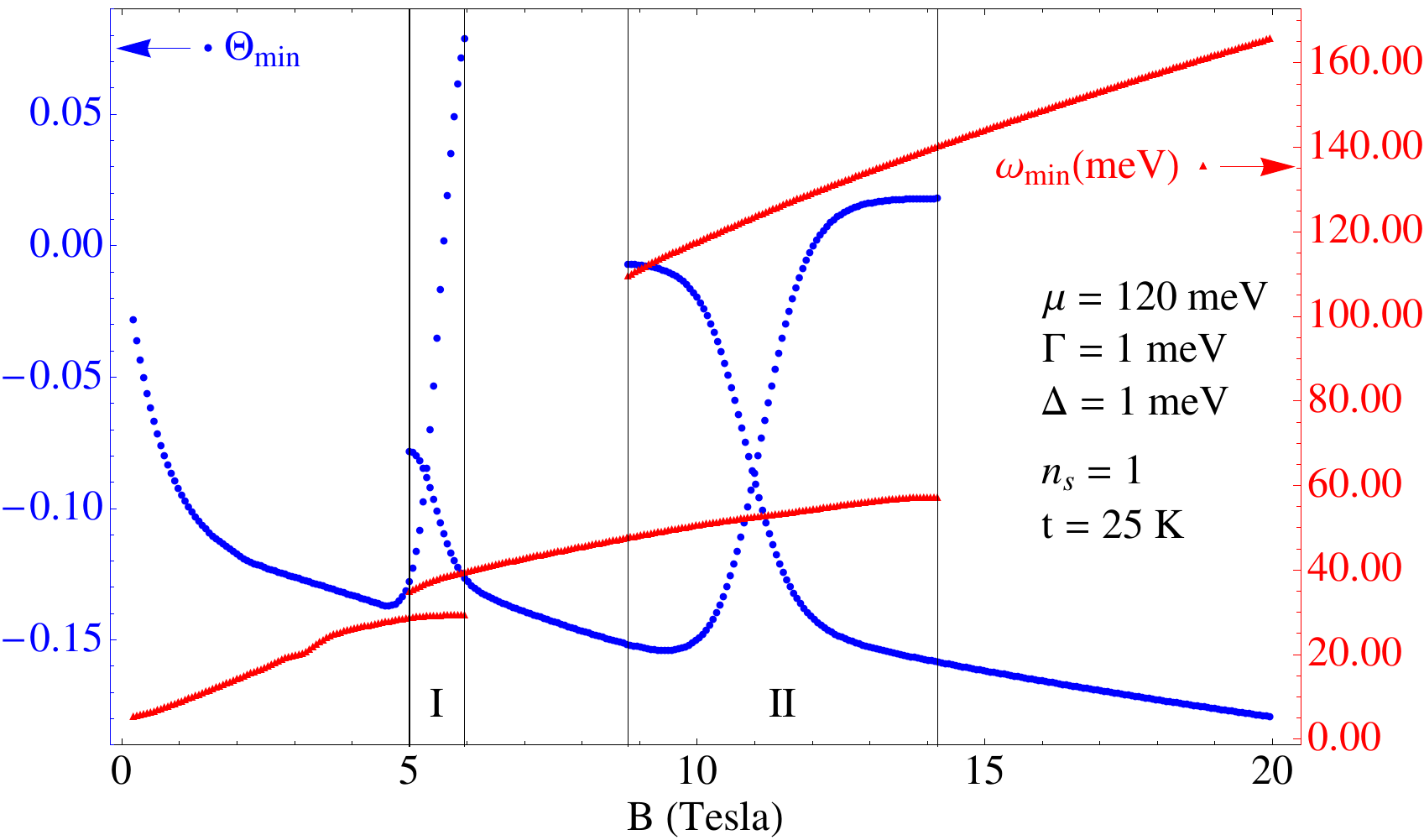} \\ 
a) & b)
\et
\caption{a) Maximum and b) minimum value of the polarization angle (left scale), and corresponding value of the light frequency (right scale), in a magnetized suspended graphene sample as a function of the applied magnetic field. The transition regions (I and II) where the shift effect manifests are the same as in Fig.\,\ref{fig4}.}
\label{fig6}
\end{center}
\end{figure}

\end{document}